# Biased exonization of transposed elements in duplicated genes: A lesson from the TIF-IA gene

Maayan Amit[1], Noa Sela[1], Hadas Keren[1], Ze'ev Melamed[1], Inna Muler[1,2,3], Noam Shomron[4], Shai Izraeli[1,2,3] and Gil Ast*[1]

Address: [1]Department of Human Molecular Genetics and Biochemistry, Sackler Faculty of Medicine, Tel-Aviv University, Ramat Aviv 69978, Israel, [2]Chaim Sheba Cancer Research Center, Tel Hashomer, Israel, [3]Pediatric Hemato-Oncology, Sheba Medical Center, Tel Hashomer, Israel and [4]Department of Biology, Massachusetts Institute of Technology, Cambridge, Massachusetts 02139, USA

Email: Maayan Amit - maayana2@post.tau.ac.il; Noa Sela - noasela@post.tau.ac.il; Hadas Keren - hadasker@post.tau.ac.il; Ze'ev Melamed - zeevmela@post.tau.ac.il; Inna Muler - imuler@yahoo.com; Noam Shomron - nshomron@mit.edu; Shai Izraeli - Shai.Izraeli@sheba.health.gov.il; Gil Ast* - gilast@post.tau.ac.il

* Corresponding author

## Abstract

**Background:** Gene duplication and exonization of intronic transposed elements are two mechanisms that enhance genomic diversity. We examined whether there is less selection against exonization of transposed elements in duplicated genes than in single-copy genes.

**Results:** Genome-wide analysis of exonization of transposed elements revealed a higher rate of exonization within duplicated genes relative to single-copy genes. The gene for TIF-IA, an RNA polymerase I transcription initiation factor, underwent a humanoid-specific triplication, all three copies of the gene are active transcriptionally, although only one copy retains the ability to generate the TIF-IA protein. Prior to TIF-IA triplication, an *Alu* element was inserted into the first intron. In one of the non-protein coding copies, this *Alu* is exonized. We identified a single point mutation leading to exonization in one of the gene duplicates. When this mutation was introduced into the TIF-IA coding copy, exonization was activated and the level of the protein-coding mRNA was reduced substantially. A very low level of exonization was detected in normal human cells. However, this exonization was abundant in most leukemia cell lines evaluated, although the genomic sequence is unchanged in these cancerous cells compared to normal cells.

**Conclusion:** The definition of the *Alu* element within the TIF-IA gene as an exon is restricted to certain types of cancers; the element is not exonized in normal human cells. These results further our understanding of the delicate interplay between gene duplication and alternative splicing and of the molecular evolutionary mechanisms leading to genetic innovations. This implies the existence of purifying selection against exonization in single copy genes, with duplicate genes free from such constrains.

# Background

The human and mouse genome sequencing projects revealed that 99% of human genes have a homolog or an ortholog in mouse, with an average of 88% conservation in the coding sequence [1]. This suggests that other factors must contribute to the phenotypic differences between human and mouse. Gene duplication and alternative splicing are two fundamental mechanisms that shape genome evolution. Alternative splicing acts at the level of mRNA processing, whereas gene duplication affects genomic DNA. Gene duplication can also operate at the level of RNA via retroposition, which has been shown to generate functional intronless duplicates of entire genes [2-5]. The contributions of these two processes to the proteome variability are substantially different [6,7].

Duplication of existing genes is an important mechanism for generating new genes while maintaining the original [8]. Gene duplication gives rise to a state of genetic redundancy, in which one of the newly formed gene copies enters a period of reduced evolutionary pressure, allowing entirely novel functional patterns to emerge. Selective constraints ensure that one of the duplicates retains its original function but the second copy is free from these constraints and, thus, accumulates mutations. These mutations can lead to a non-functional pseudogene that may continue (during a transition period) to be expressed at the RNA level; eventually the pseudogene accumulates further mutations that inactivate its transcription [9]. Alternatively, the mutations may lead to a different expression pattern or to neo-functionalization that advances organism speciation [10]. Neo-functionalization of duplicated genes was previously attributed to amino acid sequence changes through sporadic mutations or through changes in gene expression patterns [11-13]. Indeed, in plants whole genome duplication is associated with speciation [12].

An average human gene is 28,000 nucleotides long and consists of 8.8 exons of ~130 nucleotides each (excluding terminal exons) that are separated by 7.8 introns [14]. RNA splicing is the process in which introns are removed and exons are joined together to form a mature mRNA. RNA splicing is carried out by the spliceosome, which is comprised of more than 150 proteins and five snRNPs, called U1, U2, U4, U5, and U6 [15]. Alternative splicing generates multiple mRNA products from a single gene, contributing to transcriptome and proteome diversity. Alternative splicing is a possible mechanism for bridging the gap between the gene count in an organism's genome and its level of phenotypic complexity [16-18]. Up-to-date estimates suggest that more than 60% of human genes undergo alternative splicing [18]. About 80% of alternative splicing events change the encoded protein and ~15% of genetic diseases are caused by mutations within splicing regulatory elements [19]. There are four types of alternative splicing alternative 5' splice or 3' splice site selection, exon skipping, and intron retention. Selection of previously un-used splice sites can result in creation of a new exon, which is alternatively spliced. Exonization is essentially a birthing process of new exons from intronic sequences.

In human, most of newly generated exons originate from the primate-specific transposed element, *Alu*. Repetitive DNA sequences are found in most organisms and, in some, constitute a substantial fraction of the entire genome (~46% in human). Various types of repetitive DNA sequences are found within mammalian genomes and have contributed to mammalian evolution [20]. The *Alu* sequences are short interspersed elements (SINE) of about 300 nucleotides in length, which are unique to primates [21-24]. Over the past 65 million years, the *Alu* sequence has been amplified via an RNA-mediated transposition process to a copy number of 1.1 million, comprising an estimated 10% of the human genome [14,24-27].

Seven hundred thousand *Alu* elements exist within intronic sequences; of these, 480,052 are found within introns of protein coding genes, in both sense and antisense orientations with respect to the mRNA precursor [28,29]. The *Alu* element in the antisense orientation contains most of the characteristics required for identification as an exon by the splicing machinery. *Alu* exonization is an evolutionary pathway that generates primate-specific transcriptomic diversity [30].

Recent studies that examined the evolutionary trend of alternative splicing after gene duplication revealed that duplicate genes have fewer alternatively spliced forms than single-copy genes (singletons) and that an inverse correlation exists between the mean number of alternative splice forms and the gene family size [31]. These results suggest that there is a loss of alternative splicing in duplicated genes after the duplication and that there is an asymmetric evolution of alternative splicing after gene duplication [32-34]. It seems that the duplication event compensates for a reduced use of alternative splicing and that gene duplication and alternative splicing do not evolve independently [31,34].

In this study, we compared the level of *de novo* exonization of transposed elements in duplicated genes with the level in singletons. We employed a whole-genome bioinformatic analysis, supported by *in vivo* analysis, to examine whether duplicated genes exhibited a lower level of selection against exonization of transposed elements (TE). The results suggest that alternatively spliced exons that originate from exonization of transposed elements are found significantly more frequently in duplicated genes than in singletons. In one of these genes, TIF-IA, the TE in one of the duplicates, but not in the original, is exonized. We identified the point mutation leading to the exonization. When this mutation was inserted into the original gene, it caused exonization that substantially reduced production of the protein-coding mRNA. This implies that there was selection against exonization in the original gene, whereas the duplicate was free from such constraints. The exonization in the duplicated gene occurs in some leukemia cells, but not in normal cells, implying that changes in activity or concentration of splicing factors [35] within leukemia cells changes exon/intron recognition.

## Results
### *Transposed elements more likely to be exonized in duplicated genes*

Duplicated genes undergo relaxation of selective pressure following duplication. The increase in mutational rate leads to changes in the mRNA sequence in the duplicated gene relative to the original [36,37]. In this study, we investigated whether the relaxation in selective pressure also leads to an increase in exonization of TEs in duplicate genes compared with singletons. For this purpose, we compiled a dataset of exons resulting from TE exonizations within the human genome [28]. These exons were divided into two groups: those that reside within duplicated genes and those that reside within singletons. Our dataset of TE exonizations contains 1824 exons in 1388 different genes. We found 57 *Alu* exons in 45 duplicated genes, 7 MIR (mammalian interspersed repeat) exons in 7 duplicated genes, 15 L1 (LINE-1) exons in 15 duplicated genes, three L2 (LINE-2) exons in three duplicated genes, 15 LTR (long terminal repeat) exons in 12 genes, and three DNA exons in three genes. Overall, we identified 100 TE exons in 77 duplicated genes, that is, an average of 1.3 TE exons per gene (the genes with numerous exonizations are listed in Table 1; also see Additional file 1). All other TE exons (1724) reside in 1559 genes, that is, an average of 1.1 exons per gene. The exonization rate within duplicated genes was significantly higher than that in singletons (two-tailed Mann-Whitney p-value < 0.005).

Are these TE exonizations within duplicated genes part of a neo-functionalization or a non-functionalization process? Apparently, both trajectories are present within our dataset. An example for neo-functionalization is the primate-specific duplication of the gene GON4L that generated the duplicated gene YY1AP1 [38,39]. The gene YY1AP1 is functional and also has a new role as a co-activator, YY1; it also has different levels of expression within human tissue relative to its duplicate [38,40]. Our exonization analysis revealed that YY1AP1 contains two different *Alu* exonizations that do not exist within GON4L. One of the exonizations results from the *Alu* that was inserted prior to the gene duplication. The other resulted from an insertion of an *Alu* element after the duplication and subsequent exonization [41].

An example of non-functionalization and subsequent pseudogenesis is found in the duplication of the NCF1 gene. NCF1 is one of the genes responsible for the chronic granulomatous disease (CGD) and also contributes to autoimmunity [42]. In human, the gene underwent triplication, the two duplicates are well characterized [43]. Our analysis of TE exonization revealed that there was an *Alu* exonization within the second intron of each of the two pseudogenes. There is no evidence for this exonization within the functional gene, even though the *Alu* element is present within the intron of the functional gene as well. Our dataset contains both exonization of TEs that were inserted prior to gene duplication and those that were inserted after gene duplication. Two of the *Alu* exonizations in duplicated genes in our dataset were shown to be associated with cancerous tissues based on the tissue origin of their ESTs [28,44]; the *Alu* exonization within intron 2 of the gene YY1AP1, the aforementioned duplicate of the gene GON4L [38]; and *Alu* exonization within intron 6 of the gene ACAD9, a paralog of ACADVL (p-value < 0.01).

**Table 1: Duplicated genes with multiple TE exonization**

| No. of exons | TE | Gene | Accession | Description | Dup. gene | Dup. accession | Dup. description |
|---|---|---|---|---|---|---|---|
| 3 | AluSx AluJo MER5A | SRPN | NM_022807 | small nuclear ribonucleo-protein polypeptide N | SNRPB | BC003530 | small nuclear ribonucleo-protein polypeptides B and B1 |
| 2 | AluSx AluJb | NCF1 pseudo-gene | BC002816 | neutrophil cytosolic factor 1 pseudo-gene | NCF1 | NM_000265 | neutrophil cytosolic factor 1 |
| 2 | AluSx AluJb | YY1AP1 | BC025272 | YY1 associated protein | GON4L | BX648802 | gon-4-like protein |
| 2 | AluSp MIR | NSUN5C | NM_148936 | NOL1/NOP2/Sun domain family member 5C | NSUN5 | NM_018044 | NOL1/NOP2/Sun domain family, member 5 |
| 2 | AluY AluJb | RAP1A | NM_001010935 | RAP1A, member of RAS oncogene family | RAP1B | NM_015646 | RAP1B, member of RAS oncogene family |
| 2 | AluSx MSTA | POLR2J2 | BC086857 | RPB11b1beta protein (DNA directed RNA polymerase II polypeptide J-related) | POLR2J2 | NM_032958 | DNA directed RNA polymerase II polypeptide |
| 2 | AluSg/x AluJb | ZNF611 | NM_030972 | zinc finger protein 611 | ZNF701 | NM_018260 | zinc finger protein 701 |
| 2 | AluSq AluSp | ZNF702 | NM_024924 | zinc finger protein 702 | ZNF137 | NM_003438 | zinc finger protein 137 |
| 3 | MSTA MSTA MSTA | PMS2L14 | AB017005 | PMS2L14 protein. | PMS2 | NM_000533 | PMS2 postmeiotic segregation increased 2 isoform |
| 2 | AluJb AluY | ZNF283 | AK098175 | zinc finger protein 283 | ZNF420 | NM_144689 | zinc finger protein 420 |
| 2 | AluJo L1MA9 | ZNF678 | NM_178549 | zinc finger protein 678 | ZNF479 | AF277624 | zinc finger protein 479 |
| 2 | AluSx LTR49 | PILRB | NM_013440 | paired immunoglobulin-like type 2 receptor beta | PVRIG | NM_024070 | poliovirus receptor related immunoglobulin |

### *TIF-IA gene underwent a hominidae-specific triplication*
One of the genes from our bioinformatic analysis is transcription initiation factor IA (TIF-IA). This transcription initiation factor directs growth-dependent regulation of RNA polymerase I. It is a 75-kDa protein and levels, or activity, fluctuate in response to cell proliferation [45]. Genetic inactivation of the transcription factor TIF-IA leads to nucleolar disruption, cell cycle arrest, and p53-mediated apoptosis [46]. Therefore, TIF-IA is a key protein in adapting cellular biosynthetic activities to cell growth [47].

The gene that encodes TIF-IA is highly conserved from yeast to human and is essential for cell survival. Alignment of the human cDNA of TIF-IA against the human genome using Blat (UCSC Genome Browser [48]) revealed two imperfect copies of TIF-IA, which presumably resulted from gene duplication. The original copy of TIF-IA gene (termed locus 15), as well as the two duplicates (termed locus 28 and locus 21), are located on chromosome 16. In addition, a processed pseudo-gene of TIF-IA is located on chromosome 2, but is not transcribed according to EST data.

The two duplicates of the gene were probably generated in a sequential manner, as illustrated in Figure 1. Non-homologous recombination was probably the mechanism for the triplication, because all the copies are on the same chromosome. In detail, the original gene from locus 15 was duplicated almost completely, except for the last exon (exon 18), to locus 28. Next, a major deletion of 3,517 bp within the duplicate, which included exons 11 and 12, occurred. Then, locus 28 was duplicated to locus 21. This duplication was also incomplete, resulting loss of the last three exons (exons 16 to 18). The deletion was confirmed experimentally (data not shown).

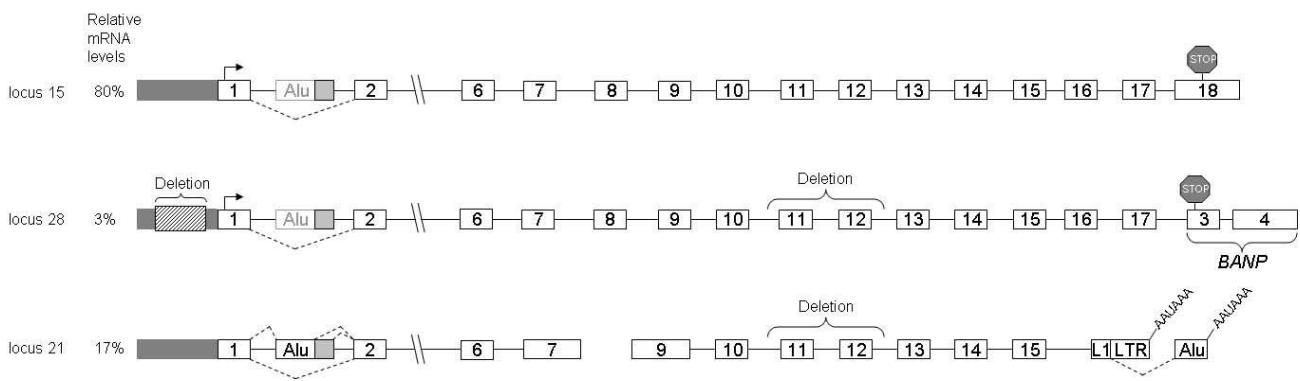

#### Figure 1
**The original TIF-IA gene and its two duplicates in the human genome**. Exons and introns are marked with numbered boxes and horizontal lines, respectively. Splicing events between exon 1 and 2 are marked with dashed lines. Each of the TIF-IA genes is named according to its relative location on chromosome 16: that is, loci 15, 28, and 21. The relative levels of expression from each locus, as measured by DS Gene®, are given as percent of total TIF-IA mRNA in P69 cells and shown on the left. Locus 15 is the original full-length gene, containing 18 exons, coding for a 75 kDa protein (651 aa) (see Additional file 3). The transcription start site is marked by an arrow above exon 1 and the stop codon by a stop sign above exon 18. The 3'UTR accounts for ~47% of the mRNA molecule. The *Alu* element is located in intron 1. Locus 28 is a duplication of the original TIF-IA gene containing the promoter region of the original gene (marked by black box). A deletion of 768 nucleotides in the promoter region reduced its transcription activity. Also, this locus contains two additional deletions with respect to the original gene: a deletion of 3517 nucleotides between intron 10 and intron 12 (between LTR16A1 and *Alu*-Sg transposed elements) and a deletion of the last exon (exon 18). The last two exons of locus 28 probably originated following a duplication event of exons 3 and 4 of BANP gene (96.6% sequence similarity). The complete BANP gene is also located on chromosome 16. The mRNA synthesized from that site has the potential to encode a protein of 514 aa according to a translation-prediction tool, using the same start codon as that of locus 15. A protein corresponding to the predicted molecular weight was not detected by western blotting analysis (see Additional file 3). Locus 21 is presumably a duplication of locus 28, containing the exact major deletion as that of locus 28 and, in addition, a deletion downstream of exon 15. The last two exons originated from exonization events of LINE-LTR and *Alu*-Sq transposed elements. Alternative poly-adenylation signal in the last two exons is shown. Locus 21 has the potential to encode a 106 aa protein from the third AUG downstream from the transcription region start site. The first AUG potentially encodes a 39-aa polypeptide. Analysis of ESTs and RT-PCR revealed that two RNA molecules are generated from locus 21, as shown in the lower panel (data not shown).

Blat of human mRNA on the chimpanzee genome suggests that it also contains more than one copy of TIF-IA gene. The original gene is located on chromosome 16, as it is in the human genome. There are several regions of homology to TIF-IA on chromosome 16 (in addition to the WT gene), as well as a processed pseudogene on chromosome 2b. It was difficult to determine the order and pattern of duplication events in the chimpanzee genome due to incomplete sequencing and the low level of genome assembly; however, it is clear that the chimp genome contains more than one copy of the TIF-IA gene on the same chromosome.

Alignment of TIF-IA cDNA with the rhesus genome revealed the presence of the original copy of the gene on chromosome 20 and the processed pseudogene on chromosome 12, but no duplications. Similar analysis on other vertebrate genomes (*C. familiaris*, *B. taurus*, *M. musculus*, *R. norvegicus*, *G. gallus*, *X. tropicalis*, *D. rerio*) and non-vertebrate organisms (*D. melanogaster*, *S. cerevisiae*, *S. pombe*, *C. elegans*) revealed that each had only a single copy of TIF-IA gene. Therefore, we concluded that the duplication of TIF-IA gene occurred after the human-chimp-rhesus split but before the human-chimp split between 4 and 25 million years ago.

### *Alu* insertion in the first intron of TIF-IA

Examination of the TIF-IA genes revealed another intriguing event: An *Alu* element was inserted into the first intron of the common ancestor of human, chimpanzee, and rhesus (approximately 25 million years ago). This *Alu* was inserted into another transposed element called L2, located in intron 1 of the TIF-IA gene. We reconstructed this scenario by examination of intron 1 of the TIF-IA gene in other mammals. The L2 element was present in all sequenced mammals except opossum. The insertion of the *Alu* element into the L2 transposed element led to an exonization in human. Based on alignment of ESTs to the human genome (see Additional file 2), the 3' splice site (3'ss) is donated by the L2 transposed element and two alternative 5' splice site (5'ss) are selected, one donated by the *Alu* sequence and the other by L2. We designated the distal 5'ss, located in the *Alu* sequence as 5'ss-a, and the proximal 5'ss, located in the L2 element, as 5'ss-b (Figure 2A and 2B). The overall steps that lead to the exonization of that exon were as follows: (1) During primate evolution, an *Alu*-Sx element was inserted into an L2 retrotransposon of the LINE family. (2) The sequence accumulated mutations that caused exonization, leading to selection of three different isoforms as demonstrated in Figure 2A. This exon is termed L2-AEx.

### Transcription and translation of the wild-type and duplicate TIF-IA genes

To examine whether the duplicates in the human genome are active transcriptionally, we examined the putative promoter regions of all three loci (Figure 1). The region of 1 kb upstream of the translation start codon of locus 15 has 96.5% identity to the corresponding upstream region of locus 21. The 1-kb promoter region of locus 28 shows similarity to that of locus 15; however, this putative promoter contains a deletion of 768 bp that ends at position -117 upstream of the potential translation start codon.

The many changes that occurred between the wild-type gene and the two duplicates suggest that the duplicates are not under selective pressure to encode functional protein. Translation of the mRNA produced from the original TIF-IA gene results in a 651-aa protein with a molecular weight of 75 kDa. The duplicate at locus 28 potentially encodes a 514-aa polypeptide chain, whereas the duplicate at locus 21 potentially encodes a 39-aa polypeptide using the genuine start codon or a 106-aa polypeptide chain using the third ATG downstream of the transcription start site (TSS). Polypeptides of sizes expected from the duplicate loci were not detected by western blotting analysis with polyclonal antibodies to the wild-type TIF-IA protein (Additional file 3). Although proteins do not appear to be generated from the duplicate genes, these duplicates are active transcriptionally as indicated by ESTs that derive from these loci. In addition, we measured the relative mRNA levels based on nucleotide variation at certain positions between the loci. In P69 cells, mRNA from the wild-type gene comprises about 80% of the mRNA transcribed from the three loci and locus 28 and locus 21 constitute 3% and 17% of the mRNA, respectively (Figure 1).

### *Alu*-exonization is exclusive to the TIF-IA duplicate in locus 21

EST analysis revealed that a large fraction of the L2-AEx-containing ESTs originated from carcinogenic tissues, such as Burkitt's lymphomas (Additional file 2). To examine the splicing patterns of exons 1 and 2 of the wild-type gene and the two duplicates, we designed a pair of primers to sequences in the flanking exons: 1 and 2 that are conserved among all three loci, meaning all three copies will be amplified simultaneously. We then performed RT-PCR analysis of human cDNA from normal tissues and from transformed cell lines. The major mRNA product from the TIF-IA genes in most of normal tissues skips the *Alu* exon and shows either no exonization or a negligible level of exonization (Figure 3A). In two cell lines, 293T and BJ-1, there was a substantial level of exonization (Figure 3A, lanes 9 and lane 10, respectively). These results imply that, for most of the tested cells, the exonization is a minor or non-existent event. We did not detect any exoni-

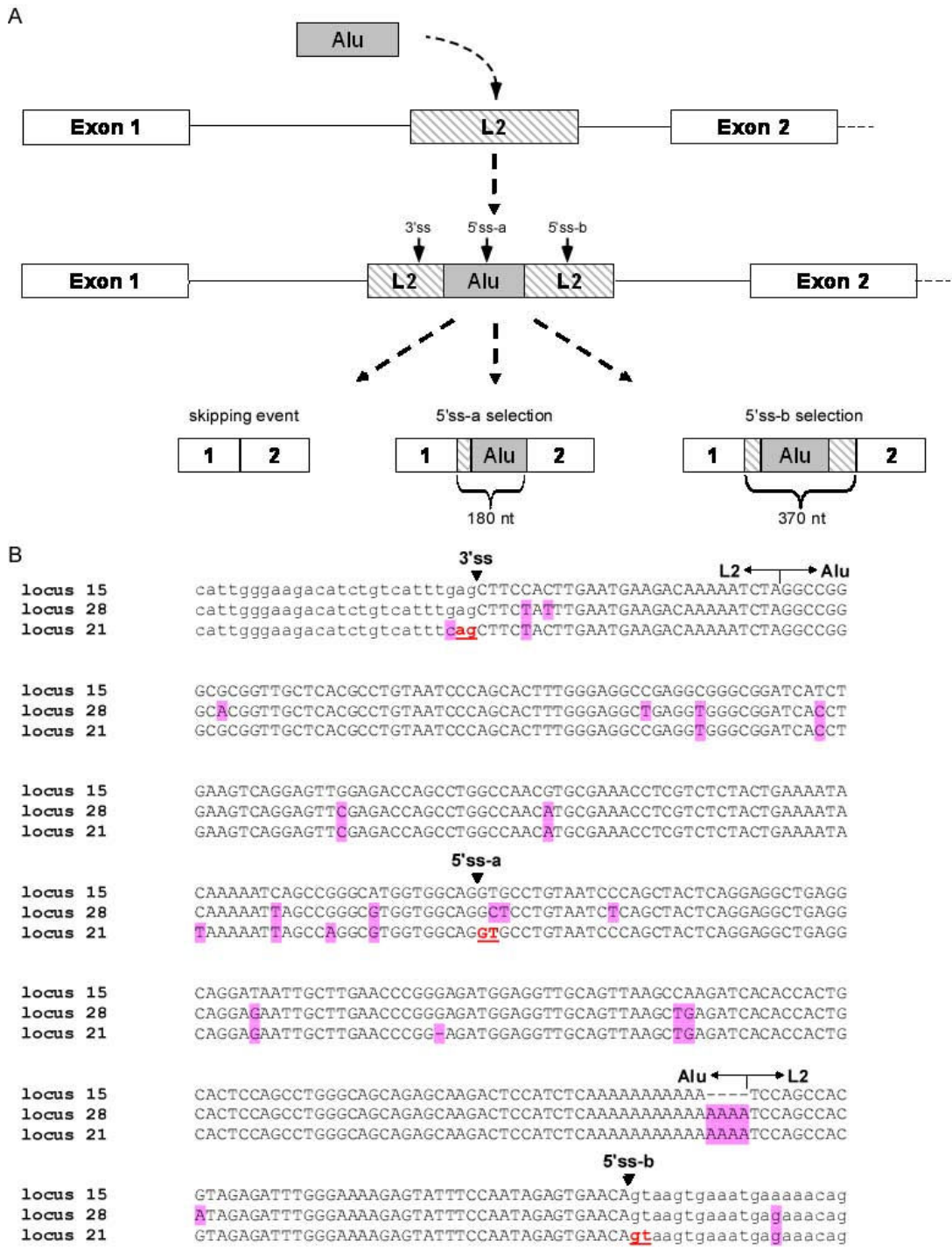

#### Figure 2
**Retrotransposition and exonization of an *Alu* element in the first intron of TIF-IA gene in locus 21**. (**A**) All mammalians genomes (except for opossum) contain a LINE element (L2) in intron 1 of TIF-IA gene. During primate evolution, an *Alu* element was inserted into L2. The L2 and *Alu* elements accumulated mutations leading to exonization (L2-AEx), in which a 3'ss and two alternative 5'ss are recognized by the splicing machinery. Three alternatively spliced isoforms are generated following this exonization: (i) a skipping isoform with no L2-AEx; (ii) selection of a distal 5'ss (termed 5'ss-a), which generates a 180-nt L2-AEx; and (iii) selection of a proximal 5'ss (termed 5'ss-b), leading to exonization of a 370-nt L2-AEx (left to right, respectively). (**B**) Multiple alignment of TIF-IA splice sites and flanking regions in all three loci. Splice sites are marked with arrows on the top. Exonic and intronic sequences are in uppercase and lower case, respectively. Mutations relative to the original gene (locus 15) are highlighted in pink. 5'ss-a is the distal splice-site and 5'ss-b is the proximal one.

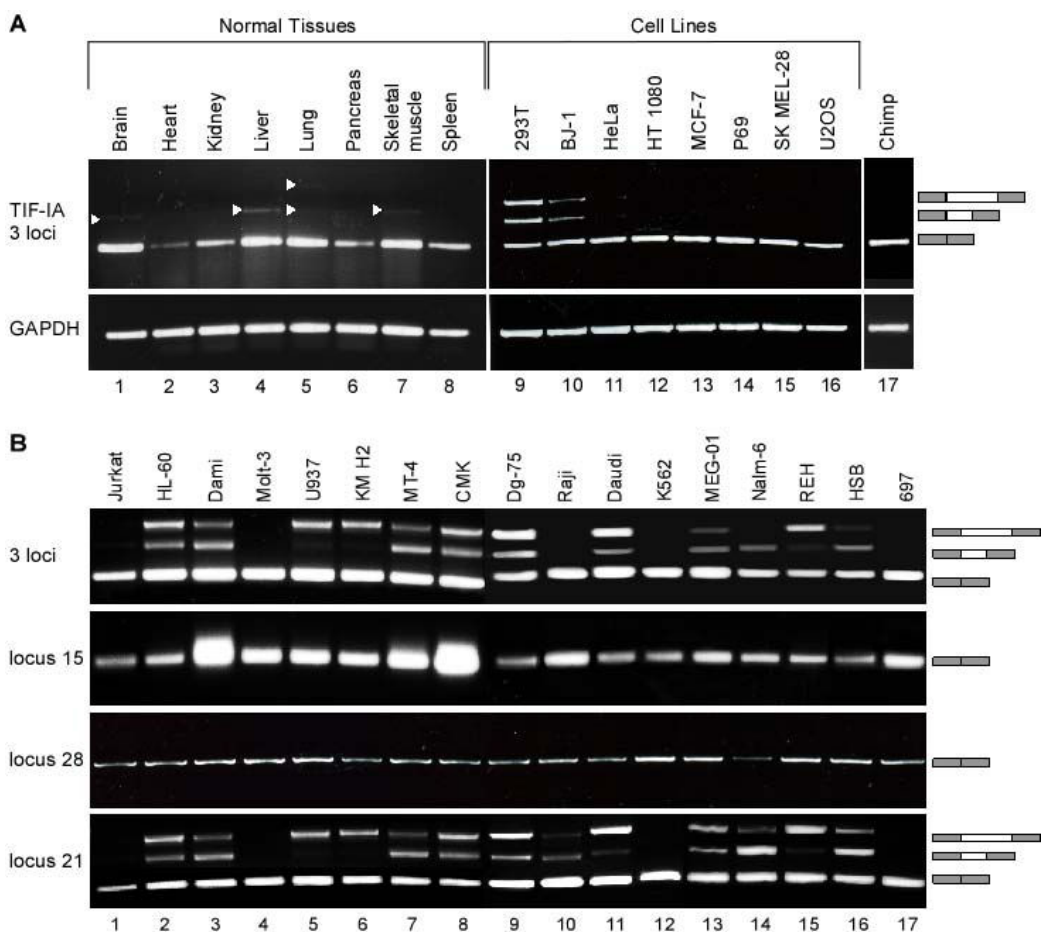

**Figure 3**
**Scanning of diverse normal human tissues, various human cell lines, and chimpanzee blood for *Alu*-exonization in the TIF-IA genes**. (**A**) Low level of L2-AEx inclusion in normal tissues. cDNA from normal human tissues and cell lines was amplified by PCR analysis with primers directed to exons 1 and 2 of all three loci of TIF-IA genes. No evidence for *Alu* exonization was observed in chimpanzee blood. Splicing products were separated on 1.5% agarose gel. The three mRNA isoforms are shown on the right; selected PCR products were eluted and sequenced. 293T are epithelial kidney cells. BJ-1 is normal human fibroblast cell line. HeLa are epithelial cervix cells with adenocarcinoma. HT 1080 originated from a fibrosarcoma. MCF-7 is a breast-cancer cell line. P69 was derived by immortalization of human primary prostate epithelial cells. SK MEL-28 is a melanoma cell line and U2OS is an osteosarcoma cell line. (**B**) Splicing patterns of L2-AEx in various human leukemia cell lines. Total cytoplasmic RNA was extracted from the indicated cell lines. Splicing products were separated on 1.5% agarose gel after RT-PCR analysis, using primers to exons 1 and 2 of all three loci of TIF-IA, to locus 15 alone, to locus 28 alone, and to locus 21 alone. The three mRNA isoforms are shown on the right. Selected splicing products were eluted from the gel and sequenced. Jurkat, Molt-3, and HSB are human T-cell leukemia cell lines. HL-60 is a myeloid cell line. Dami is a megakaryocytic AMKL (acute megakaryoblastic leukemia) non-Down syndrome (DS) cell line. U937 is derived from a human histiocyic lymphoma. MT-4 is a human T-cell lymphoblast line. Dg-75, Raji, and Daudi are Burkitt's lymphoma cell lines. KM H2 is a human Hodgkin's lymphoma cell line. CMK is a megakaryocytic DS. K562 is a chronic myeloid leukemia (CMK) cell line. MEG-01 is a megakaryoblastic cell line. 697 is pre-B ALL cell line. Nalm-6 and REH are human precursor leukemia cell lines.

zation of L2-AEx within chimpanzee blood cells, implying that this exonization may be human specific (Figure 3A, lane 17).

Next, we examined the exonization in 17 leukemia cell lines by RT-PCR analysis and detected a high level of exonization of both the short and long exonized form in 13 out of the 17 cell lines (76%). To specifically identify the locus from which this exonization occurred, we designed loci-specific primers and repeated the RT-PCR analysis using the pair of primers that detects all three loci and the locus-specific pairs of primers (Figure 3B). Only locus 21 showed exonization in which 5'ss-a or of 5'ss-b or both were selected and the ratio among the three isoforms varied significantly among the cell lines. Thus, there was a low level or an absence of exonization in normal human tissues and in epithelial cancer cell lines. In contrast, most leukemia cell lines exhibited a high level of exonization. The exonization was restricted to one of the duplicates (locus 21), but not to the original gene (locus 15). The cells that show exonization do not exhibit a known common characteristic, such as a specific hematopoietic lineage or mutations in a certain pathway or gene that can explain the exonization in these cells but not in the others. These results imply that there is a certain level of transcriptomic instability of the TIF-IA locus in certain cancer cells, in particular leukemic. The definition of exon/intron is abnormal for the TIF-IA locus, and perhaps other loci as well, in these cancer cells; what is defined as an intron for normal tissues is selected as an exon in leukemic cells.

### The wild-type TIF-IA gene is one step away from exonization

To understand the molecular mechanisms leading to exonization in locus 21, but not in loci 15 and 28, we compared the genomic sequence of the L2-*Alu* among the loci (Figure 2B). Many point mutations have accumulated in the L2-*Alu* genomic sequence following the duplication event. One of these mutations was of particular interest, because it changed a GAG found in locus 15 and 28 to CAG in locus 21. This CAG is used as the 3'ss of the L2-AEx. GAG is a week 3'ss, whereas CAG is a strong one [49]. We, thus, hypothesized that this mutation leads to locus 21-specific exonization.

To examine this hypothesis, we cloned a mini-gene containing the human genomic DNA from exon 1 to exon 2 of locus 15 (WT gene). The mini-gene was transfected into human 293T cells, RNA was extracted, and the splicing pattern between exon 1 and 2 was examined by RT-PCR analysis using primers that are specific to the mini-gene mRNA. As expected the WT mini-gene does not exhibit *Alu*-exonization (Figure 4B, lane 2). However, a single point mutation at position -3 (G → C) at the putative 3'ss activated the exonization (Figure 4B, lane 3). The predom-inant mRNA generated following this mutation is the one that includes the L2-AExs (selection of 5'ss-b), leading to a low level of the normal mRNA. In addition, the 3'ss mutation also generated an intron retention isoform. This indicates that the L2-*Alu* in the wild-type gene (locus 15) is one mutation away from exonization. If such a mutation were to occur, it would terminate production of a normal TIF-IA protein in the cells almost completely, because the L2-*AEx* inserts a premature stop codon.

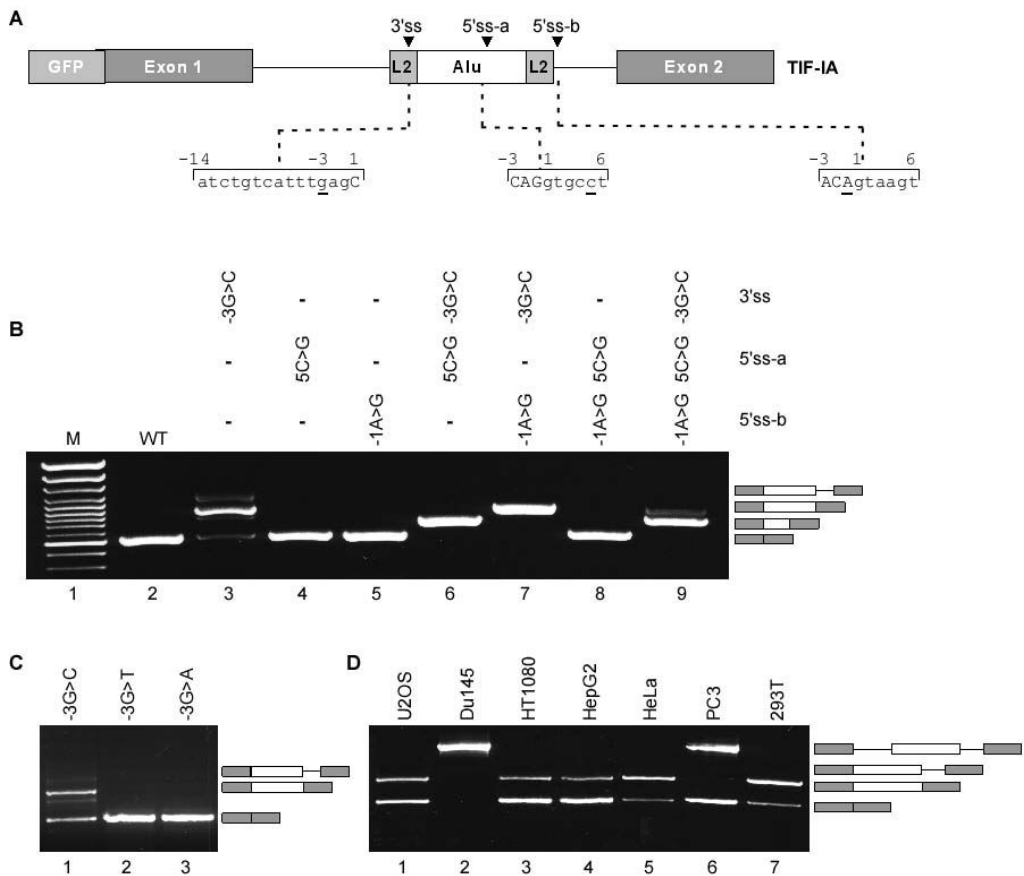

**Figure 4**
**A single point mutation activates the L2-AEx in wild-type TIF-IA**. (**A**) A schematic description of the TIF-IA mini-gene cloned into pEGFP-C3 vector, which contains the human genomic sequence from exon 1 to 2 of locus 15. The sequences of the splice sites are shown below. (**B**) The WT mini-gene and the indicated mutants were transfected into 293T cells. Cytoplasmic RNA was extracted and splicing products were separated on 1.5% agarose gel after RT-PCR analysis, using primers specific for the mini-gene RNA. The mRNA isoforms are shown on the right; the difference between the two upper products is due to alternative 5' splicing of the L2-AEx. (**C**) Similar analysis to panel B. Position -3 of the 3'ss was mutated from G to each of the other three nucleotides. (**D**) Different selection of the L2-AEx among cells. Transfection of TIF-IA mini-gene with a G → C mutation at position -3 of the 3'ss of the L2-AEx to seven different cell lines (the name of each cell line is indicated above the lane). The four splicing products are illustrated on the right. From bottom to top are: L2-AEx skipped isoform, selection of 5'ss-b *Alu*-containing exon, intron retention isoform, and unspliced mRNA. U2OS is a human-bone-osteosarcoma epithelial cell line. Du145 is a prostate-cancer cell line. HT1080 is a fibrosarcoma cell line. HepG2 is a hepatoma cell line. HeLa cells are human epithelial cells from a fatal cervical carcinoma. PC3 is a prostate cancer cell line and 293T is a human-embryonic-kidney cell line.

We next evaluated the effect of 5'ss strength on exonization. Strengthening of the either of the two sites did not activate the exonization (Figure 4B, lanes 4 and 5) and neither did strengthening both (Figure 4B, lane 8). When the 3'ss is functional, the selection of 5'ss-a or 5'ss-b is determined by their relative strength (Figure 4B, lanes 6, 7, and 9). Overall, these results indicate that the L2-*Alu* element in the WT TIF-IA gene is on the edge of exonization and that a single point mutation from G to C in position -3 of the putative 3'ss leads to exonization.

Next, we examined which nucleotides in position -3 of the 3'ss would activate exonization. We mutated the 3'ss from GAG to CAG, TAG, and AAG (Figure 4C, lanes 1, 2, and 3, respectively). Only the mutation from G to C at position -3 lead to exonization: The mutations to A and T did not. These results support our previous observation that CAG is the strongest 3'ss [49]. Also, the results show that alternative splicing is delicately balanced and is partially controlled by the strength of the 3' and 5'ss signals.

Alternative splicing is often regulated in a tissue-specific or developmental-stage-specific manner. A common explanation for differential splicing patterns is that the concentrations of splicing regulatory proteins vary depending on tissue type and developmental stage. Therefore, we examined the effect of different cellular environments on the splicing of the TIF-IA mini-gene by transfection into different cell lines; we used the mini-gene containing the G-to-C mutant at position -3 of the 3'ss. The patterns of splicing observed depended on the cell line (Figure 4D). In two cell lines, Du145 and PC3, the introns were not always excised (Figure 4D, lanes 2 and 6, respectively; see figure legend for the source of each cell line) and in U2OS, HT1080, HepG2, HeLa, and 293T cell lines, the alternatively spliced mRNA containing L2-AEx was the predominant isoform (lanes 1, 3, 4, 5, and 7, respectively). These findings show that, for this TIF-IA mini-gene, the cellular environment regulates the level of exonization and intron/exon recognition.

### *Alu-exonization in YY1AP, the duplicate of GON4Lb*
To further support the bioinformatics analysis we demonstrated another case of a gene that underwent a duplication event, and the duplicate gene exhibits a distinct alternative splicing pattern originated from an intronic *Alu* element, which also exist in the ancestor gene, but is not exonized. Figure 5 demonstrates the genomic structure of the original GON4Lb gene compared to the duplicated gene, YY1AP1 [38]. An intronic *Alu* element is exonized between exons 13 and 14 (Fig. 5A). Aligning the *Alu* element in both original and duplicated genes uncovers many sequence changes between them (Fig. 5B). In contrast to the TIF-IA gene, no mutation was detected in the splice signals, thus we assume that the reason for exonization in the duplicated gene and not in the original GON4Lb gene is due to mutations in regulatory sequences such as ESRs [50]. Figure 5C examines the splicing pattern of the corresponding genomic region and revealed that the *Alu* element in GON4Lb gene is not exonized while the duplicated gene (YY1AP1) exhibits low levels of exonization. When the *Alu*-exon enters into the mature RNA of YY1AP1, it inserts a pre-mature termination codon (PTC) [51]. This strengthened our bioinformatics analysis showing a second example of exonization that is found in the duplicated gene and not in the original copy.

### Discussion
Gene duplication and alternative splicing are two mechanisms that enhance genome and transcriptome complexity. Gene duplication works at the level of DNA and alternative splicing at the level of mRNA. These two processes are seemingly independent, but recent comprehensive bioinformatics analyses suggest that they are correlated inversely [7,31]. That is, certain genes proliferate and acquire new functions by duplication, while other genes gain new functional properties through alternative splicing.

It was observed previously that duplicated genes, unlike singletons, undergo relaxation in selective pressure. The duplicate rapidly diverges from the original sequence due to a higher rate of nucleotide substitutions within the coding regions relative to orthologs [9,36]. Here we show that there is another contributor to the fast divergence of duplicated genes: a higher level of exonization of transposed elements in duplicated genes relative to singletons. Previous work dealing with the correlation between alternative splicing and gene duplications indicated that there is less alternative splicing in duplicated genes and that there is alternative splicing loss after duplication [31,34]. All together our observations imply that although in general there is a reduction in alternative splicing in gene families, the level of TEs exonization is higher in duplicated genes.

In the analysis described here, we found that, following duplication, genes can acquire new alternatively spliced isoforms through exonization of transposed elements and that genes with more than one copy per genome (dupli-

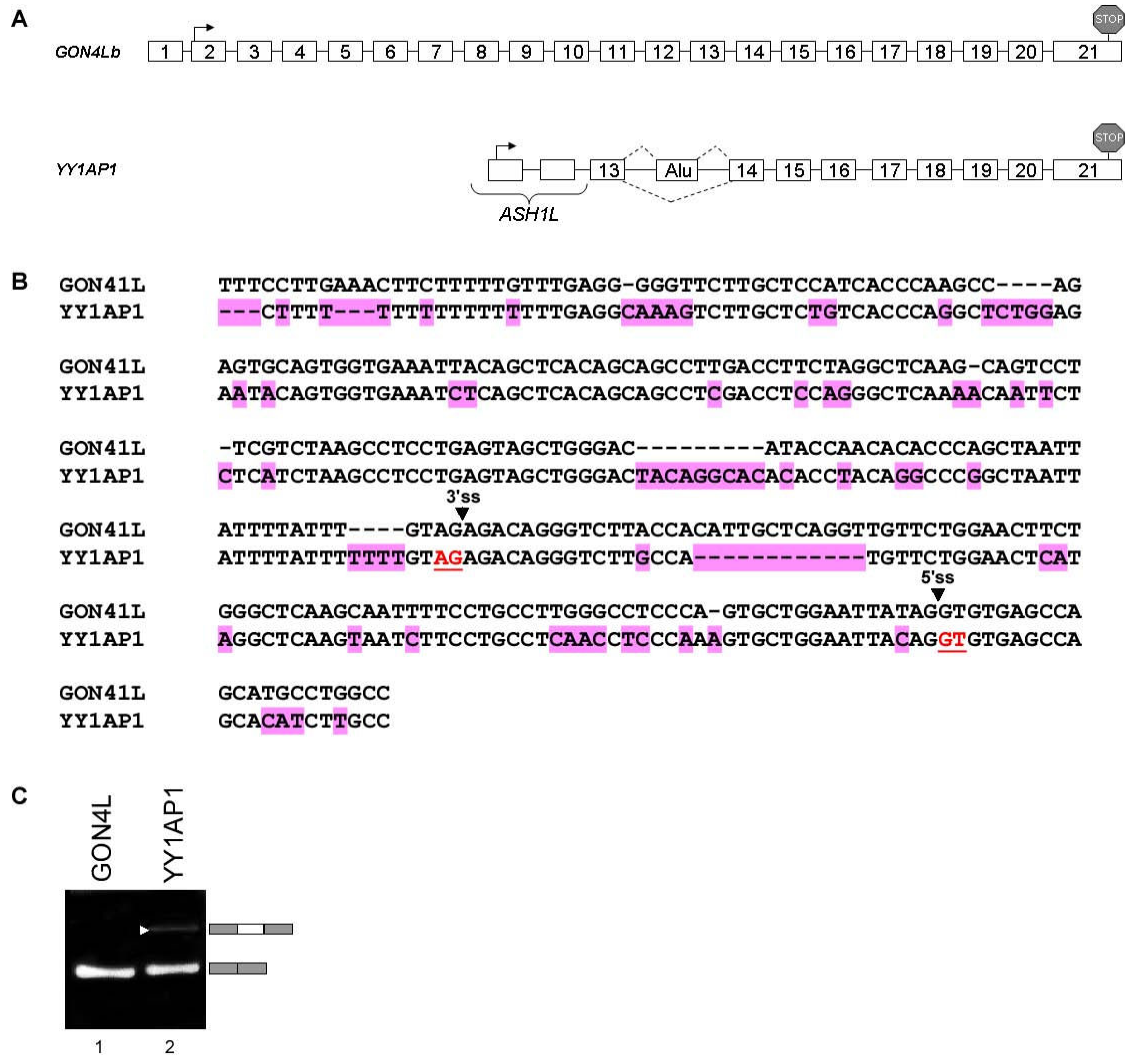

### Figure 5
**TheGON4Lb gene and its duplicate YY1AP1**. (**A**) Schematic representation of the genomic organization of GON4Lb and YY1AP1. Start codons and stop codons are marked with black arrows and stop signs, respectively. Insertion of an *Alu*-exon to the transcript of YY1AP1 insert a premature termination codon (PTC) (**B**) Multiple alignment of the alternative *Alu*-exon of YY1AP1 and the relative intronic region of GON4L. Splice sites are marked with black arrows. Substitutions and deletions relative to the original gene are marked in pink. (**C**) Exonization of an *Alu* exon in YY1AP1, and not in the original gene, as shown by RT-PCR analysis of U2OS cell line. Primers were designed to the flanking exons of each copy.

cated genes) tend to undergo more exonization of TE elements in one of the duplicates than do singletons. This indicates a high level of selection against exonization in singletons, with fewer restrictions against exonizations in multi-copy genes. It also supports the observation that one of the duplicates is under lower selective pressure, which allows the accumulation of mutations leading to exonization without deleterious effects on the organism.

Our results suggest that alternatively spliced exons that originate from exonization of transposed elements are found significantly more frequently in duplicated genes than in singletons. We selected one of these duplicated genes, TIF-IA, for further examination. TIF-IA underwent a humanoid-specific triplication, in which one of the duplicates retained the evolutionary conserved exon-intron structure and maintained the reading frame. Because the two duplicates were under less evolutionary constraint, each underwent substantial changes with respect to the original copy. These changes include large and small deletions and point mutations in the original coding sequence. In addition, in one of the duplicates, we identi-fied an exonization of an *Alu* element that was inserted into the first intron prior to the duplication. All three copies of TIF-IA gene are active transcriptionally, but the wild-type gene is the major contributor to TIF-IA mRNA levels. Only the wild-type mRNA is translated into protein. We found negligible levels of this exonization in diverse normal tissues and diverse cell lines, with the exception of the 293T and BJ-1 cell lines (Figure 3A). In contrast, when we examined various leukemia cell lines, we discovered high levels of L2-*Alu*-exonization in 13 of the 17 cell lines. This indicates that the definition of this *Alu* as an exon is restricted to certain types of cancers. The genomic sequence of the *Alu* between the cells that show exonization and those that do not is unchanged (not shown). This implies that this exonization is due to aberrant expression or activity of one or more splicing factors.

Next, we generated a mini-gene with the genomic region of locus 15, the original copy of TIF-IA gene. The single product of this wild-type mini-gene was the one that joined exon 1 with exon 2; the *Alu* element was intronic. Introducing a single point mutation at position -3 of the 3'ss (from G to C) in the *Alu* element strengthened the 3'ss [49]. The main spliced isoform in this mutant was the one in which 5'ss-b was selected. Alternative splicing of *Alu* exons usually generates both the isoform containing the additional exon and the original isoform that presumably holds an important function in the cell. Thus, alternative exonization with low inclusion level can advance transcriptomic diversification. We previously demonstrated that alternatively spliced *Alu* exons have low inclusion levels (10–19%) [52]. This strongly suggests that a point mutation that strengthened the *Alu* element 3'ss were to occur in the original gene of TIF-IA (locus 15), there would be a substantial decrease in the amount of the evolutionary conserved protein. Under these conditions, a genetic disease might develop [46].

TIF-IA (locus 15) is a vital gene [46], however its two duplicates (loci 28 and 21) are probably non-functional based on their relative short open reading frames and very low expression levels. This duplication probably represents the non-functional stage described by Ohno in the period of time following the duplication event. As described by Ohno, this initial phase of non-functionalization will be followed by complete relaxation of selective constraints and may be followed by a neo-functionalization process, wherein the duplicate acquires a new function [8,53]. The NCF1 gene duplicates may also be in the non-functional stage as its two well-described pseudogenes are presumably non-functional [43].

Why might there be selective forces against acquisition of the exons in certain genes and others not? We have demonstrated recently that conserved human-mouse alternative exons that disrupt the reading frame (termed non-symmetrical) tend to undergo fixation in the beginning of the gene, whereas those in the middle and the 3' half of the genes were presumably "weeded out" during evolution [51]. This suggests that acquisition of new exons close to the beginning of the genes is tolerated. In contrast, those in other parts of the genes are deleterious because polypeptides translated from these mRNAs are less likely to be targeted for the nonsense mediated decay than shorter products and the exon-containing isoforms are likely to be deleterious to the cells. If the initial exonization is natural or even present beneficial advantages to the cells, it will lead to an increase in the inclusion level of these alternative exons [54]. The mutation in the 3'ss of the L2-AEx in TIF-IA gene (locus 15) presumably occurred during evolution in some individuals, but was selected against since TIF-IA is a vital gene, this isoform must have been deleterious due to the insertion of a premature termination codon.

## Conclusion
Our results add another layer to our understanding of the delicate interplay between gene duplication and alternative splicing. Both increase gene complexity, albeit through different mechanisms. We have demonstrated an indirect link between duplication of genes and exonization of transposed elements. Duplicated genes, especially the non-functional, tend to undergo exonization more than singletons.

## Methods
### Dataset of TE exons in human and mouse genomes
We compiled a dataset of TE exonizations within protein coding genes [28]. The human NCBI 35 (hg17, May 2004) and the mouse NCBI33m (mm6, March 2005) assemblies were downloaded, along with their annotations, from the UCSC genome browser database [55]. Coordinates of the EST and cDNA mapping were obtained from chrN_intronEST and chrN_mrna tables, respectively. The TE mapping was obtained from chrN_rmsk tables. A TE was considered intragenic if there was no overlap with ESTs or cDNA alignments; it was considered intronic if it was found within an alignment of an EST or cDNA within an intronic region. Finally, a TE was considered exonic if it was found within an exonic part of the EST or cDNA, if it possessed canonical splice sites, and if it was an internal exon of the EST/cDNA found within the CDS or a UTR. The insertions of TEs within EST/cDNA alignments were separated into two parts: those that entered within protein coding genes relative to the list of the knownGene table in the UCSC genome browser [55] (based on proteins from SWISS-PROT, TrEMBL, TrEMBL-NEW and their corresponding mRNA from GenBank), and other insertions within cDNA/ESTs alignments that were not mapped to the known genes list and, therefore, were considered as non-protein-coding genes. Non-protein-coding genes were determined as genomic regions covered by at least two correctly spliced cDNA/ESTs (flanked by canonical splice sites) containing at least three exons that did not overlap any annotated gene based on UCSC known genes lists, versions hg17 and mm6 for human and mouse, respectively. Unspliced genes were not included in our analysis; we only considered genes with at least two introns. Internal UTR exons were considered to be internal according to the annotations of knownGenes in UCSC and the fact that they were internal within the cDNA/EST. The TE position within the gene (UTR or CDS) and the exon phase were calculated based on knownGenes table annotations of the gene start and end positions, as well as CDS start and end positions.

### TE exonization within duplicated genes
To analyze whether the exonization occurred within a duplicated gene or a singletons, we performed a blast search of these genes against all mRNA sequences listed as known genes (knownGene table) downloaded from UCSC human genome build hg17 [55] and searched for transcripts with at least 75% sequence similarity along 40 percent of the gene, the dataset was filtered to delete transcripts of isoforms generated by alternative splicing that belonged to the same gene and nonduplicated genes. The filtration of similar mRNA that are isoforms of the same gene was done by comparing the locus of the mRNA as indicated in UCSC genome browser; only mRNAs that mapped to different loci were considered.

### Cell line maintenance
293T, HeLa, HT1080, HepG2, Du145, and U2OS cell lines were cultured in Dulbecco's Modification of Eagle Medium (DMEM), supplemented with 4.5 g/mL glucose (Biological Industries, Inc., Israel), 10% fetal calf serum (FCS), 100 U/mL penicillin, 0.1 mg/mL streptomycin, and 1 U/mL nystatin (Biological Industries, Inc.). PC3 cells were cultured in Ham's F12K medium with 2 mM L-glutamine adjusted to contain 1.5 g/L sodium bicarbonate (90%) and fetal bovine serum (10%). Cells were grown in 6-well plates or 100-mm culture dishes under standard conditions at 37°C with 5% $CO_2$. Cells were split at a 1:10 ratio twice weekly.

### Transfection
Cells were grown to 50% confluence in 100-mm culture dishes or 6-well plates and maintained as described above. Twenty-four hours prior to transfection, cells were split, and transfection was performed using FuGENE6 (Roche) with 0.5–1 μg of plasmid DNA. Cells were harvested after 48 hr.

### RNA isolation and splicing analysis
Total RNA was extracted using TRI Reagent (Sigma), followed by treatment with 2 U of RNase-free DNase (Ambion). Reverse transcription (RT) was performed on 1–2 μg total RNA in a 12.5 μL final volume containing: 100 mM DTT, 10 mM dNTPs, 100 mM oligo(dT) primer, 2 U of reverse transcriptase avian myeloblastosis virus (RT-AMV, Roche), and RT buffer. The final mixture was then incubated for 1 hr at 42°C. The spliced cDNA products derived from the expressed mini-genes were detected by PCR, using Taq polymerase (BioTools), and pEGFP-C3 specific reverse and forward primers. Primer sets for cell-line scanning purposes were designed to amplify all loci in a single PCR or to amplify every locus separately: 3 loci: 5'-CGT TAG TTC GGC CCA ATG-3', 5'-CTT CAG CAA GAC TTC TGT CAC-3'; locus 15: 5'-CTT CGT CCT CTG CAG TTA AGA AG-3', 5'-CTT CAG CAA GAC TTC TGT CAC-3'; locus 28: 5'-CGC TTC GCC CTC TGC AGT C-3', 5'-CTT CAG CAA GAC TTC TGT CAC-3' after reverse transcription (RT) with unique primer; locus 21: 5'-CTT CAC ACG TTG TTT GTC G-3', 5'-CTT CAG CAA GAC TTC TGT CAC-3'. Amplification of the chimpanzee cDNA (proliferating blood cells of a female chimpanzee from the safari in Ramat-Gan) was performed with 5'-CGT TAG TTC GGC CCA ATG-3' and 5'-GCT GGT TCT TCA ACA ACT CAA A-3'. The primers flank the *Alu* element in intron 1 of the TIF-IA's genes. Human GAPDH: 5'-TCG TGG AGT CCA CTG GCG TCT T-3' and 5'-TGG CAG TGA TGG CAT GGA CTG T-3'. Chimp GAPDH: 5'-TCG TGG AGT CCA CTG GCG TCT T-3' and 5'-TGG CAG TGA TGG CAT GGA CCG T-3'. GON4L: 5'-ATG AGC TGA TGG AAG AGC TG-3' and 5'-GAG GGG TGT TAA AGT TAG GAC GAG-3'. YY1AP1: 5'-CAA ATG AGC TGA TGG AAG AT-3' and 5'-GAG GGG TGT TAA AGT TAG CTT-3'. Amplification was performed for 30 cycles, consisting of denaturation for 30 seconds at 94°C, annealing for 45 seconds at 58°C, and elongation for 1.5 minutes at 72°C. The products were separated in 1.5% agarose gel; selected bands were confirmed by sequencing.

### Plasmid construct
A genomic DNA from 293T cell line (Gentra) corresponding to exon 1 through exon 2 of the TIF-IA gene (from locus 15) was PCR amplified and cloned into pEGFP-C3 vector (Clontech) between XhoI and BamHI sites under the control of the human cytomegalovirus (CMV) immediate early promoter, giving a ~1.9 kb insert. F: 5'-AAA AAA ACT CGA GGC TGA TTG GCT GAA GGT TG-3'; R: 5'-AAA AGG ATC CCA GCA ATA GTT GTA TTC TGA CCT AAC C-3'.

### Site-directed mutagenesis
Specific overlapping oligonucleotide primers that contained the desired mutation were used to insert the mutation using PfuTurbo DNA polymerase (Stratagene). After

PCR amplification, the reaction was digested with DpnI restriction enzyme (New England Biolabs) for 1 hr at 37°C; 1–3 μL of the reaction were transformed into *E. coli* XL-1 strain and colonies were picked for mini-prep extraction (Qiagen) and sequenced. L2-AEx-3'ss (-3)G->C: 5'-GAA GAC ATC TGT CAT TTC AGC TTC CAC TTG AAT G-3' and 5'-CAT TCA AGT GGA AGC TGA AAT GAC AGA TGT CTT C-3'. L2-AEx-3'ss (-3)G->T: 5'-GAA GAC ATC TGT CAT TTT AGC TTC CAC TTG AAT G-3' and 5'-CAT TCA AGT GGA AGC TAA AAT GAC AGA TGT CTT C-3'. L2-AEx-3'ss (-3)G->A: 5'-GAA GAC ATC TGT CAT TTA AGC TTC CAC TTG AAT G-3' and 5'-CAT TCA AGT GGA AGC TTA AAT GAC AGA TGT CTT C-3'. L2-AEx-5'ss-a (5)C->G: 5'-CAT GGT GGC AGG TGC GTG TAA TCC CAG CTA C-3' and 5'-GTA GCT GGG ATT ACA CGC ACC TGC CAC CAT G-3'. L2-AEx-5'ss-b (-1)A->G: 5'-GTA TTT CCA ATA GAG TGA ACG GTA AGT GAA ATG AAA AAC AGC-3' and 5'-GCT GTT TTT CAT TTC ACT TAC CGT TCA CTC TAT TGG AAA TAC-3'.

### Western blotting
Lysis buffer (50 mM Tris, pH 7.5; 1% NP40; 150 mM NaCl; 0.1% SDS; 0.5% deoxycholic acid; protease inhibitor cocktail and phosphatase inhibitor cocktail I and II; Sigma) was used for protein extraction. Then lysates were centrifuged for 30 minutes at 14,000 rpm at 4°C. Total protein concentrations were measured using BioRad Protein Assay (BioRad). Proteins were separated in 10% SDS-polyacrylamide gel electrophoresis (SDS-PAGE) and then electroblotted onto a Protran membrane (Schleicher and Schuell). The membranes were probed with polyclonal anti-TIF-IA antibody (kindly provided by Ingrid Grummt) at 1/10000 dilution followed by rabbit secondary antibody. Immunoblots were visualized by enhanced chemiluminescence (Lumi-Light Western Blotting Substrate; Roche) and exposure to x-ray film.

### Transposed elements analysis
RepeatMasker® software version 3.1.0 [56] was used for the detection of transposed elements.

### Analysis of the relative mRNA levels
The PCR product from RT-PCR of P69 cells was excised and purified following electrophoresis on 1.5% agarose gel (Promega, Madison, WI, USA). Direct sequencing was performed using the ABI PRISM (Applied Biosystems, Foster-City, CA, USA). The variation percentage from direct sequencing was calculated for the reverse primer; the presented percentages represent an average of three separated positions (31, 63 and 105) along exon 2 of the mRNAs. The nucleotides were quantified by the Discovery Studio (DS) Gene 1.5 program (Accelrys Inc., San Diego, CA, USA).

### List of abbreviations
TE, transposed elements; 3'ss, 3' splice site; 5'ss, 5' splice site; SINE, short interspersed elements

### Authors' contributions
MA was responsible for analyzing different cell lines, generating the constructs, transfection experiments and drafting the manuscript. NSE executed the bioinformatic data analysis and drafted the manuscript. HK carried out the identification of TIF-IA protein. ZM performed the GON4L and YY1AP1 analysis. IM provided many of the cell culture samples. NSH participated in the cell lines characterization. SI was involved in designing the study and helped to draft the manuscript. GA conceived and supervised the study design, and drafted the manuscript. All the authors read and approved the final manuscript.

### Additional material

**Additional file 1**
*Transposed elements exonizations within duplicated genes in the human genome*. A list of transposed elements exonizations within duplicated genes and their genome coordinates.
Click here for file
[http://www.biomedcentral.com/content/supplementary/1471-2199-8-109-S1.xls]

**Additional file 2**
*Indications of exonization only from locus 21*. Alignment of EST/cDNA with the human genome using UCSC Genome Browser reveals alternative L2-AExs between exon 1 and 2. The L2-AExs' are marked with red circles. Exons' numbers are indicated on the bottom. Seven out of the 19 ESTs that cover that locus contain an L2-AEx, of which 5 were originated from cancerous sources (3 from Burkitt's lymphoma, 1 from neuroblastoma cell line, and 1 from poorly differentiated adenocarcenoma). The rest of the Alu-containing spliced isoforms originated from reproductive organs (testis and uterus). The alignment indicates that the L2-AEx is also alternatively 5' spliced.
Click here for file
[http://www.biomedcentral.com/content/supplementary/1471-2199-8-109-S2.tiff]

**Additional file 3**
*The WT TIF-IA protein is synthesized in cells with and without the exonization*. Total protein was extracted from the indicated cells and analyzed by Western blotting using polyclonal rabbit anti-TIF-IA serum (kindly gifted from the Grummt's lab) at a 1/10,000 dilution. 293T and U2OS are human cell lines, and 3T3 is a murine cell line.
Click here for file
[http://www.biomedcentral.com/content/supplementary/1471-2199-8-109-S3.tiff]

### Acknowledgements
We would like to thank Ingrid Grummt from the German Cancer Research Center (DKFZ) for providing us with the antibody against TIF-IA. We thank Eddo Kim and Oren Ram for helpful discussions and Nurit Paz for calculating the relative mRNA levels with the DS gene program. This work was supported by ICA through the Ber-Lehmsdorf Memorial Fund, and TAU Cancer Center, by the Cooperation Program in Cancer Research of the Deutsches Krebsforschungszentrum (DKFZ) and Israeli's Ministry of Science and Technology (MOST), by a grant from the Israel Science Foundation (1449/04 and 40/05), MOP Germany-Israel, GIF, N.S. is funded in part by EURASNET.